\documentclass[twoside,12pt]{article}
\usepackage{epsfig}
\usepackage{subfigure,graphicx}
\usepackage{wrapfig,rotating}
\usepackage{amssymb}
\usepackage{amsmath}

\newcommand{\be}{\begin{equation}}
\newcommand{\ee}{\end{equation}}

\newcommand\epsfigure[4][width=\hsize]{%
\begin{figure}[htbp]%
  \begin{center}%
     \IfFileExists{#2.eps.bb}%
       {\includegraphics[draft,#1]{#2}}%
       {\includegraphics[#1]{#2}}%
  \end{center}%
\caption{#3}\label{#4}%
\end{figure}%
}

\begin{document}
\title{\vspace{1cm} 
Generalized parton distributions in the context of HERA measurements
}
\author{Laurent SCHOEFFEL \\
CEA Saclay/Irfu-SPP, 91191 Gif-sur-Yvette, France
}
\maketitle
\begin{abstract}
\noindent
{\it
We present the  recent experimental results from
deeply virtual Compton scattering (DVCS) from H1 and ZEUS experiments at HERA.
Their interpretation encoded in the generalized parton distributions is discussed.

}
\end{abstract}

%
%
\section{  Introduction }

Measurements of the deep-inelastic scattering (DIS) of leptons and nucleons, $e+p\to e+X$,
allow the extraction of Parton Distribution Functions (PDFs) which describe
the longitudinal momentum carried by the quarks, anti-quarks and gluons that
make up the fast-moving nucleons. 
While PDFs provide crucial input to
perturbative Quantum Chromodynamic (QCD) calculations of processes involving
hadrons, they do not provide a complete picture of the partonic structure of
nucleons. 
In particular, PDFs contain neither information on the
correlations between partons nor on their transverse motion.
Hard exclusive processes, in  which the
nucleon remains intact, have emerged in recent years as prime candidates to complement
this essentially one dimentional picture. 
The simplest exclusive process is the deeply virtual
Compton scattering (DVCS) or exclusive production of real photon, 
$e + p \rightarrow e + \gamma + p$.
This process is of particular interest as it has both a clear
experimental signature and is calculable in perturbative QCD. 
The DVCS reaction can be regarded as the elastic scattering of the
virtual photon off the proton via a colourless exchange, producing a 
real photon in the final state  \cite{dvcsh1,dvcszeus,hermes,jlab}. 
In the Bjorken scaling 
regime, 
QCD calculations assume that the exchange involves two partons, having
different longitudinal and transverse momenta, in a colourless
configuration. These unequal momenta or skewing are a consequence of the mass
difference between the incoming virtual photon and the outgoing real
photon. This skewedness effect can
 be interpreted in the context of generalised
parton distributions (GPDs) \cite{gpds}. 

\begin{figure}[htbp]
\begin{center}
\includegraphics[width=8.cm]{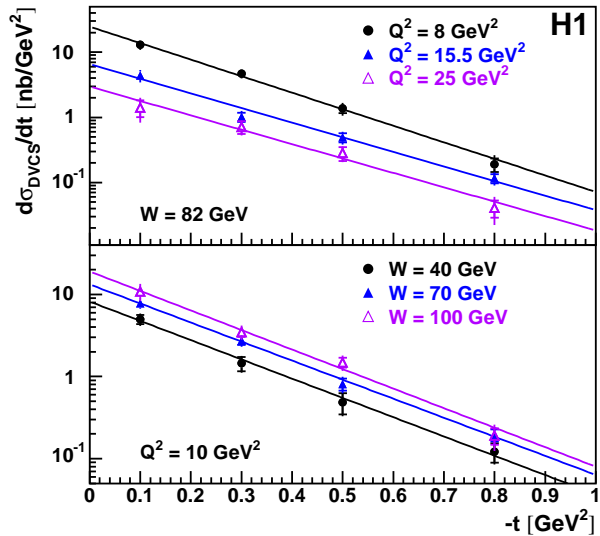}
\end{center}
\vspace*{-0.5cm}
\caption{\label{figbdvcs} 
The DVCS cross section, differential in $t$, 
for three values of $Q^2$ expressed atand 
for three values of $W$.
The solid lines represent the results of fits 
of the form $e^{-b|t|}$.
}
\end{figure}

With $t=(p-p')^2$, the momentum transfer (squared) at the proton vertex, the measurement of the 
VM and DVCS cross section, differential in $t$
 is one of the key measurement in exclusive processes.
A parameterization in $d\sigma/dt \sim e^{-b|t|}$, as shown in
 Fig. \ref{figbdvcs}, 
gives a  very good description of measurements.
In addition, in  Fig. \ref{figbdvcs}, we show that 
fits of the form $d\sigma/dt \sim e^{-b|t|}$ can
describe DVCS measurements to a very good accuracy
for different $Q^2$ and $W$ values.

Then, we can define a
generalised gluon distribution $F_g$ which depends both on $x$ and $t$
(at given $Q^2$).
From this function, we can compute
a gluon density which also depends on a spatial degree of freedom, 
a transverse size (or impact parameter), labeled $R_\perp$,
in the proton. Both functions are related by a Fourier transform 
$$
g (x, R_\perp; Q^2) 
\;\; \equiv \;\; \int \frac{d^2 \Delta_\perp}{(2 \pi)^2}
\; e^{i ({\Delta}_\perp {R_\perp})}
\; F_g (x, t = -{\Delta}_\perp^2; Q^2).
$$

From the Fourier transform relation above, 
the average impact parameter (squared), $\langle r_T^2 \rangle$,
of the distribution of gluons 
$g(x, R_\perp)$ 
is given  by
\begin{equation}
\langle r_T^2 \rangle
\;\; \equiv \;\; \frac{\int d^2 R_\perp \; g(x, R_\perp) \; R_\perp^2}
{\int d^2 R_\perp \; g(x, R_\perp)} 
\;\; = \;\; 4 \; \frac{\partial}{\partial t}
\left[ \frac{F_g (x, t)}{F_g (x, 0)} \right]_{t = 0} = 2 b,
\label{myequation}
\end{equation}
where $b$ is the exponential $t$-slope.
In this expression, $\sqrt{\langle r_T^2 \rangle}$
is  the transverse distance between the struck
parton and the center of momentum of the proton. 
The latter is the average transverse
position of the partons in the proton with weights given by 
the parton momentum fractions.
At low $x_{Bj}$, the transverse
distance defined as $\sqrt{\langle r_T^2 \rangle}$
corresponds also to the relative transverse distance
between the interacting parton (gluon in the 
equation above) and the system defined by spectator partons.
Therefore provides a natural estimate of the transverse
extension of the gluons probed during the hard process.
In other words,
a Fourier transform of momentum
to impact parameter space readily shows that the $t$-slope $b$ is related to the
typical transverse distance in the proton.
This $t$-slope, $b$, corresponds exactly to the slope measured once
the component of the probe itself contributing to $b$ can be
neglected, which means at high scale: $Q^2$ or $M_{VM}^2$.
Indeed,
at high scale, the $q\bar{q}$ dipole is almost
point-like, and the $t$ dependence of the cross section is given by 
the transverse extension 
of the gluons in the  proton for a given $x_{Bj}$ range.

DVCS results lead to $\sqrt{r_T^2} = 0.65 \pm 0.02$~fm at large scale $Q^2 > 8$ GeV$^2$ 
for $x_{Bj} \simeq 10^{-3}$ \cite{dvcsh1}.
This is not useless to recall that this observation
 is extremely challenging on the experimental
analysis side. We are dealing with nano-barn cross sections, that we measure
as a function of $t$, and finally, we measure the energy dependence of this 
behavior in $t$. 
Of course, the gain is  important.
In particular, the great interest of the DVCS is that the $t$ dependence
measured is free of effects that could come from VM wave functions (in case of VMs)
and then spoil (to a certain limit) 
the interpretation of $b$ described above.
Thus,
with DVCS, we have the advantage to work in a  controlled environment
(photon wave functions)
where the generic Eq. (\ref{myequation}) can be applied  to
the measurement (almost directly) and must not be
corrected with effects arising from VMs wave function.

%
%
\section{  Generalised parton distributions}

In the previous section, we have shown that
data on DVCS can give access to the
spatial distribution of quarks and gluons in the proton
at femto-meter scale. 
Then, we have defined functions, which model this property (for gluons)
through the relation
$$
g (x, R_\perp; Q^2) 
\;\; \equiv \;\; \int \frac{d^2 \Delta_\perp}{(2 \pi)^2}
\; e^{i ({\Delta}_\perp {R_\perp})}
\; F_g (x, t = -{\Delta}_\perp^2; Q^2).
$$
Of course, a similar relation holds for quarks, linking
the two functions
$q(x, R_\perp; Q^2) $ and $ F_q (x, t = -{\Delta}_\perp^2; Q^2)$.
The general framework for this physics is encoded in the  so-called
generalized parton distributions (GPDs) (see Ref. \cite{gpds,imaging}).

We already know that the reconstruction of spatial images from scattering
experiments by way of Fourier transform of the observed 
scattering pattern is a technique widely used in physics,
for example, in X-rays scattering from crystals.
In simple words, what we have done experimentally is that
we have extended this technique
to the spatial distribution of quarks and gluons
within the proton, using processes that probe the proton at a tiny resolution scale.
Of course, as already mentioned, working at a femto-meter scale
with nano-barn cross sections
is very challenging from the experimental front. We have achieved this
and it immediately opens a way in the ambitious program of
mapping out the GPDs.
We come back below in a more systematic way on different
aspects of that program that requires
a large amount of experimental informations, for which future programs at JLab
and CERN are appealing.

\begin{figure}[htbp]
\begin{center}
  \includegraphics[width=.4\textwidth]{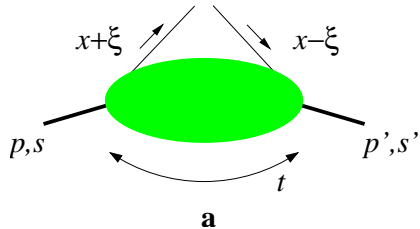}
    \caption{  Picture of a GPD and its variables.  The
  momentum fractions $x$ and $\xi$ refer to the average hadron
  momentum $\frac{1}{2} (p+p')$. Note that $x$ is an internal variable
  and is not equal to $x_{Bj}$. However,
  there is a relation between the skewing variable $\xi$ and 
  $x_{Bj}$, $\xi = x_{Bj}/(2-x_{Bj})$.}
\label{fig:gpd}
\end{center}
\end{figure}

Let us give a short overview of GPDs,
in simple terms. It is interesting, even for 
an experimentalist, as it clarifies the Fourier 
transform relation discussed above and makes
more transparent the goals for the future.
For complete reviews, see Ref. \cite{gpds,imaging}.
GPDs are defined through matrix elements $\langle p' | \mathcal{O} | p
\rangle$ between hadron states $|p'\rangle$ and $|p \rangle$, with
non-local operators $\mathcal{O}$ constructed from quark and gluon
fields.  
From this expression, we understand why GPDs are directly related
to the amplitude for VM or real gamma exclusive production.  
For unpolarized quarks there are two
distributions $H^q(x,\xi,t)$ and $E^q(x,\xi,t)$, where $x$ and
$\xi$ are defined in Fig. \ref{fig:gpd}.  The former is
diagonal in the proton helicity, whereas the latter describes proton
helicity flip.  For $p=p'$ and equal proton helicities, we recover
the diagonal matrix element parameterized by usual quark and antiquark
densities, so that $H^q(x,0,0)=q(x)$ and $H^q(-x,0,0)=-\bar{q}(x)$ for
$x>0$.  
Note that the functions of type $E$ are not accessible in 
standard DIS, as it corresponds to matrix elements
$\langle p',s' | \mathcal{O} | p,s
\rangle$ with $s \ne s'$. Even in DVCS-like analysis, it is 
very difficult to get a sensitivity to these functions,
as in most observables, their contributions are 
damped by kinematic factors of orders $|t|/M_p^2$,
with an average $|t|$ value in general much smaller that $1$ GeV$^2$.
Then, till stated otherwise, our next experimental discussions
are concentrated on the determination of GPDs of type
$H_q$ or $H_g$.
We come back later on this point and show specific cases
where $E$-type functions can be accessed and why this is
an important perspective.

An interesting property of GPDs, which lightens their
physics content, is that 
their lowest moments give the well-known Dirac and Pauli
form factors
\begin{equation}
\sum_q e_q \int dx\, H^q(x,\xi,t) = F_1(t) 
\qquad\qquad
\sum_q e_q \int dx\, E^q(x,\xi,t) = F_2(t),
\label{ffactors}
\end{equation}
where $e_q$ denotes the fractional quark charge. 
It means that GPDs measure the contribution of quarks/gluons,
with longitudinal momentum fraction $x$, to the corresponding 
form factor. In other words,  GPDs are
like mini-form factors that filter out quark with a 
longitudinal momentum fraction $x$
in the proton.
Therefore, in the same way as Fourier transform of a 
form factor gives the charge
distribution in position space, Fourier transform of GPDs (with
respect to variable $t$)
contains information about the spatial distribution of partons in the proton.

The strong interest in determining GPDs of type $E$
is that these functions appear in a 
fundamental relation between GPDs and angular momenta of partons.
Indeed,
GPDs have been shown to be related directly to the total 
angular momenta carried by partons in the nucleon, via the Ji relation~\cite{gpds}
\begin{equation}
\frac{1}{2} \int_{-1}^1 dx x\left(H_q(x,\xi,t) + E_q(x,\xi,t)\right) = J_q .
\label{eq:JiRelation}
\end{equation} 
As GPDs of type $E$ are essentially unknown apart
from basic sum rules, any improvement of their knowledge is essential.
From Eq. (\ref{eq:JiRelation}), it is clear that we could access directly to the
orbital momentum of quarks if we had a good knowledge of GPDs $H$ and $E$.
Indeed,  $J_q$ is the sum of the longitudinal angular momenta of quarks
and their orbital angular momenta. The first one is relatively well known
through global fits of polarized structure functions.
It follows that a determination of $J_q$ can provide an estimate of
the orbital part of its expression. 
In Ji relation (Eq. (\ref{eq:JiRelation})), the function
$H$ is not a problem as we can take its limit at $\xi=0$,
where $H$ merges with the PDFs, which are well known.
But we need definitely to get a better understanding of $E$.

In order to give more intuitive content to the Ji relation (\ref{eq:JiRelation}),
we can comment further its dependence in the function $E$.
From our short presentation of GPDs, we know that
functions of type $E$ are 
related to matrix elements of the form
$\langle p',s' | \mathcal{O} | p,s
\rangle$ for $s \ne s'$, which means helicity flip at the proton vertex ($s \ne s'$).
That's why their contribution vanish in standard DIS or in processes where
$t$ tends to zero. More generally, their contribution would vanish
if the proton had only configurations where
helicities of the partons add up to the helicity of the proton.
In practice, this is not the case due to angular momentum of partons.
This is what is reflected in a very condensed way in the Ji relation (Eq. (\ref{eq:JiRelation})).

Then, we get the intuitive interpretation of this formula: it connects
$E$ with the angular momentum of quarks in the proton.
A similar relation holds for gluons \cite{gpds}, linking $J_g$
to $H_g$ and $E_g$ and
both formulae, for quarks and gluons, add up to build the proton spin  
$$
J_q+J_g = 1/2.
$$
This last equality must be put in perspective with the
asymptotic limits for $J_q$ and $J_g$ at large scale $Q^2$,
which read $J_q \rightarrow \frac{1}{2} \frac{3 n_f}{16+3n_f}$ and
$J_g \rightarrow \frac{1}{2} \frac{16}{16+3n_f}$, 
where $n_f$ is the number of active flavors of quarks
at that scale (typically $n_f=5$ at large scale $Q^2$) \cite{gpds}.

In words, half of the angular momentum of the proton is carried by gluons
(asymptotically). It is not trivial  to make quantitative estimates
at medium scales, but it is a clear indication that orbital angular
momentum plays a major role in building the angular momentum of the proton.
It implies that all experimental physics issues
that intend to access directly or indirectly to GPDs of type $E$
are essential in the understanding of the proton structure,
beyond what is relatively well known concerning 
its longitudinal momentum structure in $x_{Bj}$.
And that's also why first transverse target-spin asymmetries
(which can provide the best sensitivity to $E$)
are so important and the fact that such measurements have already
been done is promising for the future \cite{gpds,imaging}.

Clearly, we understand at this level the major interest of GPDs
and we get a better intuition on their physics content.
They simultaneously probe
the transverse and the longitudinal distribution of quarks and gluons
in a hadron state and the possibility to flip helicity in GPDs
  makes these functions sensitive to orbital angular momentum
in an essential way.
This is possible because they generalize the
purely collinear kinematics describing the familiar twist-two
quantities of the parton model. This is obviously
illustrating a fundamental 
feature of non-forward exclusive processes \cite{gpds,imaging}.

\section{  Outlook }

We have reviewed the most recent experimental results from
DVCS at HERA. Exclusive processes in DIS, like DVCS, have appeared
as key reactions to trigger the generic mechanism of diffractive scattering.
Decisive measurements have been performed recently, which 
provide first  experimental features concerning proton 
tomography, on how partons are localized in the proton.
A unified picture of this physics is encoded in the GPDs formalism.




\begin{thebibliography}{99}

\bibitem{dvcsh1}
C.~Adloff {\it et al.}  [H1 Collaboration],
Phys.\ Lett.\  B {\bf 517} (2001) 47;  \\ 
%
A.~Aktas {\it et al.}  [H1 Collaboration],
Eur.\ Phys.\ J.\  C {\bf 44} (2005) 1;  \\ 
%
F.~D.~Aaron {\it et al.}  [H1 Collaboration],
Phys.\ Lett.\  B {\bf 659} (2008) 796; 
arXiv:0907.5289 [hep-ex].

\bibitem{dvcszeus}
P.~R.~B.~Saull  [ZEUS Collaboration],
arXiv:hep-ex/0003030; \\  
%
S.~Chekanov {\it et al.}  [ZEUS Collaboration],
Phys.\ Lett.\  B {\bf 573} (2003) 46; 
  JHEP {\bf 0905} (2009) 108.



\bibitem{hermes}
A.~Airapetian {\it et al.}  [HERMES Collaboration],
Phys.\ Rev.\ Lett.\  {\bf 87} (2001) 182001; 
Phys.\ Rev.\  D {\bf 75} (2007) 011103;  
JHEP {\bf 0806} (2008) 066.


\bibitem{jlab}
S.~Stepanyan {\it et al.}  [CLAS Collaboration],
Phys.\ Rev.\ Lett.\  {\bf 87} (2001) 182002; \\ 
%
C.~Munoz Camacho {\it et al.}  
[Jefferson Lab Hall A Collaboration and Hall
A DVCS Collaboration],
Phys.\ Rev.\ Lett.\  {\bf 97} (2006) 262002; \\  
%
S.~Chen {\it et al.}  [CLAS Collaboration],
Phys.\ Rev.\ Lett.\  {\bf 97} (2006) 072002; \\  
%
F.~X.~Girod {\it et al.}  [CLAS Collaboration],
Phys.\ Rev.\ Lett.\  {\bf 100} (2008) 162002.  


\bibitem{gpds}
X.~D.~Ji,
Phys. Rev. Lett. {\bf 78} (1997) 610; Phys. Rev. {\bf D55} (1997) 7114; \\ 
%
M.~Diehl, T.~Gousset, B.~Pire and J.~P.~Ralston,
Phys.\ Lett.\  B {\bf 411} (1997) 193;  \\  
%
L.~L.~Frankfurt, A.~Freund and M.~Strikman,
Phys.\ Rev.\  D {\bf 58} (1998) 114001
[Erratum-ibid.\  D {\bf 59} (1999) 119901]; \\   
%
A.~G.~Shuvaev, K.~J.~Golec-Biernat, A.~D.~Martin and M.~G.~Ryskin,
Phys.\ Rev.\  D {\bf 60} (1999) 014015; \\     
%
K.~Goeke {et al.},
{Prog.\ Part.\ Nucl.\ Phys.}\  {\bf 47}, 401 (2001); \\     
%
A.~V.~Belitsky, D.~Mueller and A.~Kirchner,
Nucl.\ Phys.\  B {\bf 629} (2002) 323; \\   
%
M.~Diehl,
{Phys.\ Rept.}\ {\bf 388}, 41 (2003);  \\  
%
A.~V.~Belitsky and A.~V.~Radyushkin,
{Phys.\ Rept.}\  {\bf 418}, 1 (2005).

\bibitem{imaging} 
M.~Diehl,
Eur.\ Phys.\ J.\ C {\bf 25} (2002) 223
[Erratum-ibid.\ C {\bf 31} (2003) 277]; \\ 
%
J.~P.~Ralston and B.~Pire,
  Phys.\ Rev.\  D {\bf 66}, 111501 (2002); \\
%
M.~Burkardt,
Int.\ J.\ Mod.\ Phys.\ A {\bf 18} (2003) 173; \\  
%
A.~V.~Belitsky, X.~d.~Ji and F.~Yuan,
Phys.\ Rev.\  D {\bf 69} (2004) 074014; \\  
%
L.~Frankfurt, M.~Strikman and C.~Weiss,
Ann.\ Rev.\ Nucl.\ Part.\ Sci.\  {\bf 55} (2005) 403.




\end{thebibliography}
\end{document}